\documentclass[aps,twocolumn,showpacs,preprintnumbers,amsmath,amssymb]{revtex4}
\usepackage{graphicx}
\usepackage{color}

\begin{document}

\title{Vector interaction strength in Polyakov--Nambu--Jona-Lasinio models from
hadron-quark phase diagrams}
\author{O. Louren\c co$^1$, M. Dutra$^1$, T. Frederico$^1$, A. Delfino$^2$ and
M. Malheiro$^1$}

\affiliation{$^1$Departamento de F\'isica, Instituto Tecnol\'ogico da
Aeron\'autica, CTA, S\~ao Jos\'e dos Campos, 12228-900, SP, Brazil\\
$^2$Instituto de F\'isica - Universidade Federal Fluminense,
Av. Litor\^ anea s/n, 24210-150 Boa Viagem, Niter\'oi RJ, Brazil}

\begin{abstract}
We estimate the vector interaction strength of the Polyakov--Nambu--Jona-Lasinio
(PNJL) parametrizations, assuming that its transition curves should be as close
as possible of the recently studied \mbox{RMF-PNJL} hadron-quark phase diagrams.
Such diagrams are obtained matching relativistic mean-field hadronic models, and
the PNJL quark ones. By using this method we found for the magnitude of the
vector interaction, often treated as a free parameter, a range of
\mbox{$7.66$ GeV$^{-2}\lesssim G_V \lesssim 16.13$ GeV$^{-2}$}, or equivalently,
\mbox{$1.52 \lesssim G_V/G_s \lesssim 3.2$}, with $G_s$ being the scalar
coupling constant of the model. These values are compatible but restricts the
range of \mbox{$4$ GeV$^{-2}\lesssim G_V \lesssim 19$ GeV$^{-2}$}, recently
obtained from lattice QCD data through a different mean-field model approach.
\end{abstract}

\maketitle

The hadron-quark phase transition is still a challenging task for both
theoretical and experimental fields. From the theoretical point of view, the
strongly interacting matter is treated by QCD. However, in the regime of low
energies, QCD is nonperturbative and still difficult to solve for intermediate
temperatures and chemical potentials, although lattice methods have been faced a
huge progress in the last years \cite{lattice}. For such regime, it is useful to
use effective models in the description of the quark matter that share the same
features of QCD. For example, the MIT bag model \cite{mit}, the NJL one
\cite{njl,buballa} and its version coupled with the Polyakov loop, named as PNJL
model~\cite{fuku1}.

In this context such models are used to construct the QCD phase diagram
\cite{fukureview}, where the different regions are
identified as those in which the chiral symmetry is broken or restored. Studies
in this direction were performed, for instance, for the linear $\sigma$ model
\cite{kapusta}, and the NJL one \cite{cratti}. The additional information about
the confined/deconfined phases can also be taken into account when the PNJL
model is used to construct the quark phase transitions
\cite{weise4,fuku2,outros}, or even when the Polyakov loop is linked with
the linear $\sigma$ model \cite{plsm}.

In a very recent study \cite{prd}, it was performed a comparison among the
hadron-quark phase diagrams generated by PNJL models, and those constructed
by matching a large class of relativistic mean-field (RMF) hadronic models with
four different parametrizations of the PNJL quark model (other studies based on
this treatment can be found in Refs.
\cite{ditoro,ciminale,debora,delfino,muller}). The results shown pointed out to
a difference between the phase transition curves due to the repulsive
interaction of the RMF models.

Based on these results, we propose the construction of a PNJL model that
minimizes this difference. This will be done by including a vector interaction
in the original PNJL structure. We name hereafter the resulting model as
\mbox{PNJLv} model, and the strength of its vector interaction will then
estimated in order to approximate the \mbox{PNJLv} transition curves to the
\mbox{RMF-PNJL} ones as much as possible.

Actually, the inclusion of vector interactions in effective quark models was
already discussed in the literature, see Ref. \cite{buballa} for a study in the
context of the NJL model. It is known that the effect of the increase of the
repulsive interaction strength in the quark matter phase diagram is to shrink
the first-order transition region as the chemical potential  $\mu$ increases.
Also the critical end-point of the transition moves to larger $\mu$ and lower
temperature $T$. Such effects, and further studies are reported for both NJL
model \cite{kashiwa,kashiwa2} and the \mbox{PNJLv} one
\cite{fuku2,kashiwa,pnjlgv,KasPRD11,ratio3}, where the vector interaction
strength is often used as a free parameter. Therefore, taking into account the
aforementioned effects, we furnish here a method to estimate this interaction
strength, based on the \mbox{RMF-PNJL} hadron-quark phase diagrams. We also
compare our results with the recent ones, based on a mean-field calculation
related to QCD lattice data, proposed in Ref. \cite{ferroni}.

Our starting point is the Lagrangian density of the two-flavor \mbox{PNJLv}
model, that reads
\begin{eqnarray}
\mathcal{L} &=& \bar{\psi}(i\gamma_\mu D^\mu - m)\psi +
G_s\left[(\bar{\psi}\psi)^2 - (\bar{\psi}\gamma_5\vec{\tau}\psi)^2
\right] \nonumber \\
&-& G_V(\bar{\psi}\gamma_\mu\psi)^2 - \mathcal{U}(T,\mu,\Phi,\Phi^*)
\end{eqnarray}
with $D^\mu=\partial^\mu - iA^\mu$ being the covariant derivative, $A^\mu$ the
gluon field, and $m$ the current quark mass. The vector interaction strength is
regulated by the parameter $G_V$, and the Polyakov loop potential is given by
$\mathcal{U}(T,\mu,\Phi,\Phi^*)$. As in Ref.~\cite{prd}, we use four different
versions of this potential, namely RRW06 \cite{weise4,weise2}, RTW05
\cite{weise1}, FUKU08 \cite{fuku2} and DS10 \cite{schramm}.

From this Lagrangian density one obtains, following the procedure used in Ref.
\cite{weise1} and taking into account the new vector term, the grand
thermodynamic potential,
\begin{eqnarray}
&\Omega& = \mathcal{U}(T,\mu,\Phi,\Phi^*) + G_s\rho_s^2
- \frac{\gamma}{2\pi^2}\int_0^{\Lambda}E(k,M)\,k^2dk \nonumber \\
&-& \frac{\gamma}{6\pi^2}\int_0^{\infty}\mathcal{F}_+\frac{k^4dk}{E(k,M)}
-G_V\rho^2,
\label{omega}
\end{eqnarray}
in the isospin symmetric system, in which $E(k,M)=(k^2+M^2)^{1/2}$, and with the
degeneracy factor given by $\gamma=12$. The constituent quark mass is
$M=m-2G_s\rho_s$, and the quark density is obtained from
$\rho=-\partial\Omega/\partial\tilde{\mu}$, where $\tilde{\mu}$ is related with
$G_V$ and $\rho$ through $\tilde{\mu}=\mu-2G_V\rho$. Note that, the quark
density can be also found by requiring that $\partial\Omega/\partial\rho=0$. The
functions $\mathcal{F}_{\pm}=\mathcal{F}_{\pm}(k,T,\tilde{\mu},\Phi,\Phi^*)$
are defined by $\mathcal{F}_{\pm}=F(k,T,\tilde{\mu},\Phi,\Phi^*)\pm
\bar{F}(k,T,\tilde{\mu},\Phi ,\Phi^*)$.

The strength of the scalar interaction ($G_s$), the vacuum integral cutoff
($\Lambda$), and $m$ are also parameters of the model. Here they are given by
$G_s=5.04$~GeV$^{-2}$, $\Lambda=651$ MeV, and $m=5.5$~MeV~\cite{weise1}.

The inclusion of the confinement information, via the Polyakov loop, affects the
statistical distributions of the \mbox{PNJLv} model in such way that the new
Fermi-Dirac functions are now given by
\begin{eqnarray}
F(k,T,\tilde{\mu},\Phi,\Phi^*) = \frac{\Phi e^{2x} + 2\Phi^*e^{x} + 1}
{3\Phi e^{2x} + 3\Phi^* e^{x} + e^{3x} + 1}
\label{fdmp}
\end{eqnarray}
and $\,\bar{F}(k,T,\tilde{\mu},\Phi,\Phi^*)=F(k,T,-\tilde{\mu},\Phi^*,\Phi)$,
with \mbox{$x=(E-\tilde{\mu})/T$}.

The quark condensate $\left<\bar{\psi}\psi\right>=\rho_s$, and the Polyakov loop
$\,\Phi\,$ are found by requiring that $\partial\Omega/\partial\rho_s =
\partial\Omega/\partial\Phi=0$, in the lowest order approximation
\cite{weise4,weise6}, that leads to $\Phi=\Phi^*$. The explicit form of the
equations of motion (EOM) of the \mbox{PNJLv} model, found by minimizing
$\Omega$ in respect to $\rho$, $\rho_s$ and $\Phi$, are:
\begin{eqnarray}
\rho &=&
\frac{\gamma}{2\pi^2}\int_0^{\infty}\mathcal{F}_-k^2dk
-\frac{\partial\mathcal{U}}{\partial\mu}, 
\label{qdens} \\
\rho_s &=& \frac{\gamma}{2\pi^2}\int_0^{\infty}\mathcal{F}_+\frac{M}{E}k^2dk 
-\frac{\gamma}{2\pi^2}\int_0^{\Lambda}\frac{M}{E}k^2dk
\label{qrhos}
\end{eqnarray}
and
\begin{eqnarray}
\frac{\partial\mathcal{U}}{\partial\Phi} &-& \frac{T\gamma}
{2\pi^2}\int_0^{\infty}(g_1+g_2)k^2dk = 0
\label{qphi}
\end{eqnarray}
where
\begin{eqnarray}
g_1 &=& g_1(k,T,\tilde{\mu},\Phi) =\frac{1+e^{-x}}{3\Phi(1+e^{-x})+e^{x}+
e^{-2x}},
\end{eqnarray}
and $g_2=g_1(k,T,-\tilde{\mu},\Phi)$. Note that these EOM are the same PNJL
ones, with the chemical potential $\mu$ shifted by the vector interaction, as
given by $\tilde\mu$. In the RMF case, such a shift also occurs in the value of
the baryonic chemical potential due to the contribution of the mean-field value
of the isoscalar vector meson field $\omega$.

It is important to remark that we are dealing with the simplest \mbox{PNJLv}
version, in which color quark condensates are not being taken into account.
Here, we are not considering any possibility of emergence of two-flavor
superconducting color (2SC), or color flavor locked (CFL) phases \cite{sc1,sc2}.

To construct the \mbox{PNJLv} transition curves varying the $G_V$ parameter, we
follow the procedure adopted by Fukushima \cite{fuku2} that uses the magnitude
of the order parameters $\rho_s$ and $\Phi$, to define the transition
temperature for each fixed $\mu$. In that work, the author uses the condition of
$\rho_s/\rho_s^{\mbox{\tiny{vac}}}=1/2$ to construct the phase diagrams. The
transition temperature at $\mu=0$ found in that case is around $T_c(\mu=0)=200$
MeV. Noteworthy to observe that in the construction of the quark phase diagrams,
the approach using fixed values for $\rho_s$ is qualitatively equivalent, at
least at moderated $\mu$ values and for $G_V\neq0$, to those based on the local
maximum of $\partial\rho_s/\partial T$ and $\partial\Phi/\partial T$, used in
Ref. \cite{weise1}. In the $G_V=0$ case, the agreement between the crossover
transition curves obtained with the two criteria is still better.

Here, we use different values of $\rho_s/\rho_s^{\mbox{\tiny{vac}}}$ for each
Polyakov potential in order to obtain a better agreement of $T_c(\mu=0)$ with
the lattice QCD results, $T_c(\mu=0)=173\pm 8$ MeV~\cite{stringent}. The adopted
values are $\rho_s/\rho_s^{\mbox{\tiny{vac}}}=0.73,0.70,0.72,0.71$, respectively
for the RRW06, RTW05, FUKU08, and DS10 parametrizations. Furthermore, we still
maintain the rescaling of the original parameters $T_0$ and $b$ of
$\mathcal{U}(T,\mu,\Phi,\Phi^*)$ to $T_0=190$ MeV (RRW06, RTW05 and DS10),
and $b=0.007\Lambda^3$ (FUKU08) also used in Ref. \cite{prd}.

The \mbox{PNJLv} diagrams constructed from the adopted method are displayed in
Figs.~\ref{figs}a-\ref{figs}d.

From Fig. \ref{figs}, one can see that there is a range of values for $G_V$, at
least in a certain temperature region, that makes the transitions constructed
from the \mbox{PNJLv} model very close to those obtained via the \mbox{RMF-PNJL}
matchings, represented by the gray bands. For these matchings, we have used a
large class of RMF hadronic models coupled to the PNJL ones in which $G_V=0$
\cite{prd}.

The overlap between the hadron-quark phase transitions provided by the
\mbox{PNJLv} and \mbox{RMF-PNJL} models, is found for $G_V^{\mbox{\tiny
MIN}}\lesssim G_V \lesssim G_V^{\mbox{\tiny MAX}}$, with $G_V^{\mbox{\tiny
MIN}}/G_s=1.60,1.52,1.60,1.54$, and $G_V^{\mbox{\tiny
MAX}}/G_s=3.20,2.74,2.86,2.98$, respectively for the RRW06, RTW05, FUKU08, and
DS10 Polyakov potentials. This give us a total range of \mbox{$7.66$
GeV$^{-2}\lesssim G_V \lesssim 16.13$ GeV$^{-2}$}. We can also define a
temperature region of better overlap between the \mbox{PNJLv} results, and the
\mbox{RMF-PNJL} bands, given by $T\lesssim T^{\mbox{\tiny MAX}}$. The
maximum temperatures are $T^{\mbox{\tiny MAX}}\approx 90,80,50,80$ MeV, for the
same aforementioned \mbox{PNJLv} models. Therefore, our findings can be useful
for instance, in the study of proto-neutron stars (PNS) that are described at
\mbox{$T\lesssim 50$~MeV}. Applications of the \mbox{RMF-PNJL} models to compact
stars, have been done recently for the PNS evolution \cite{gshao}, for quark
\cite{manuel}, and hybrid stars \cite{jaziel}.

\onecolumngrid
\textcolor{white}{asdfas}
\vspace{0.2cm}
\begin{figure}[!htb]
\centering
\includegraphics[scale=0.26]{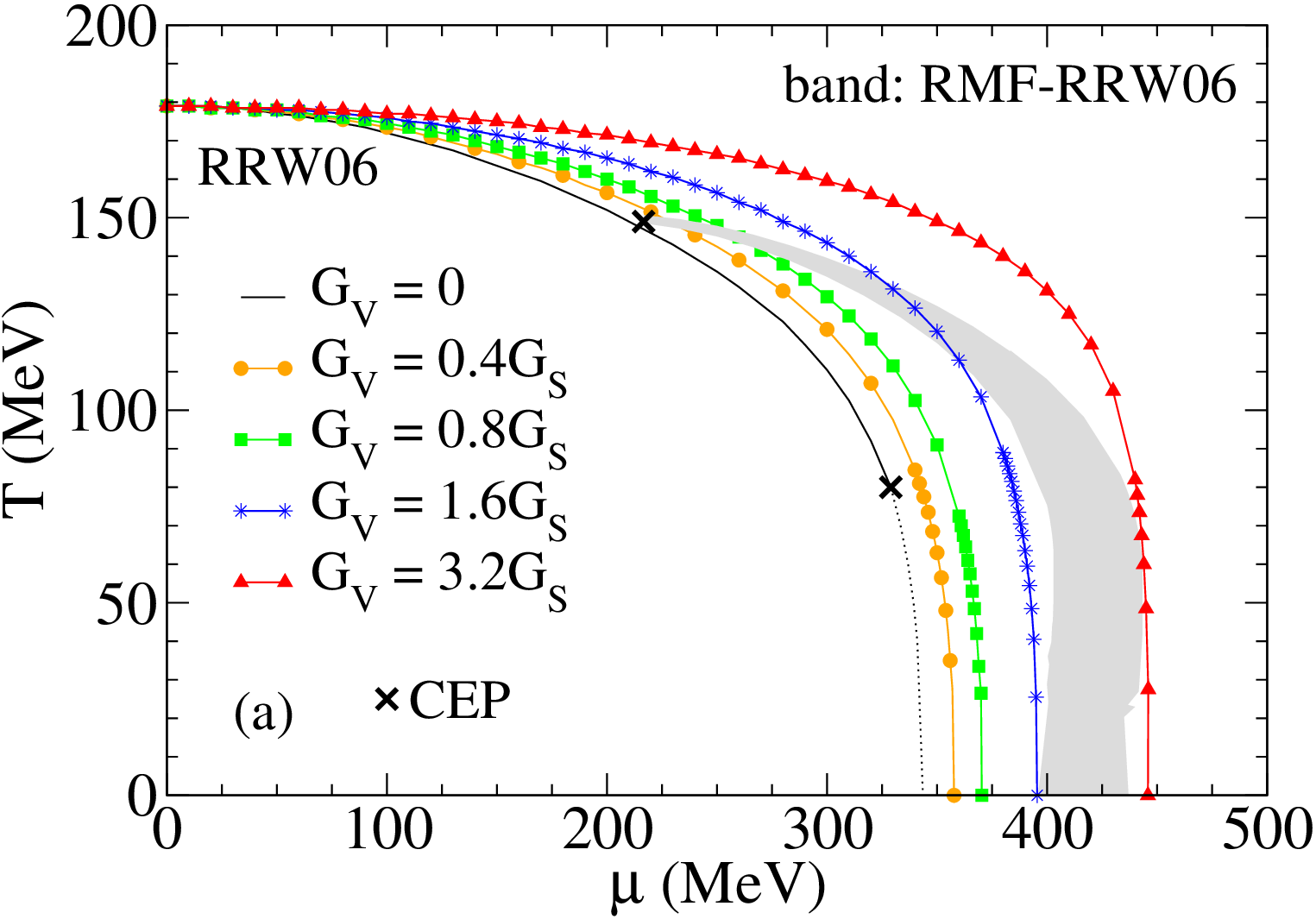}
\includegraphics[scale=0.26]{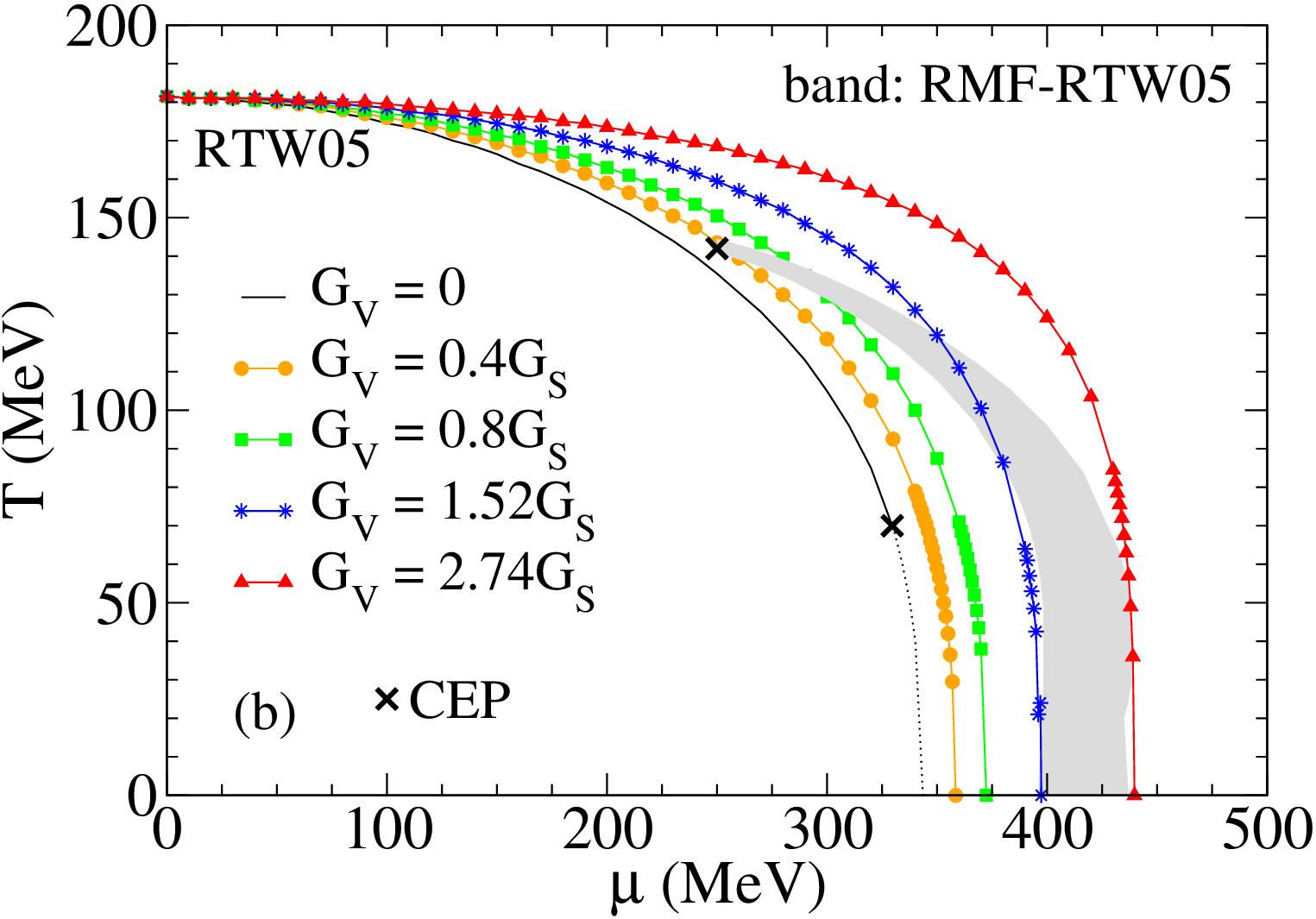}
\newpage
\end{figure}
\vspace{-0.9cm}
\begin{figure}[!htb]
\centering
\includegraphics[scale=0.26]{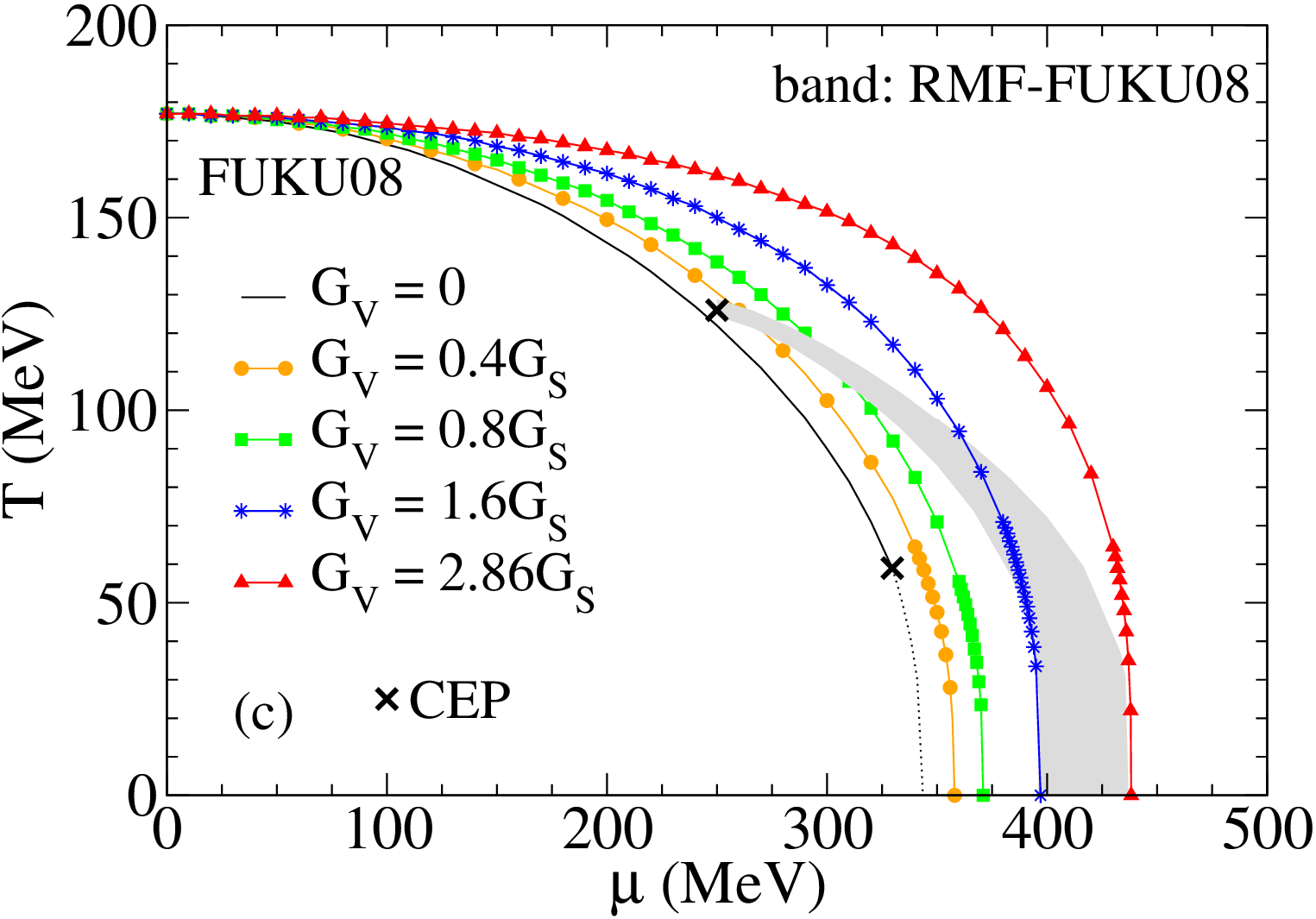}
\includegraphics[scale=0.26]{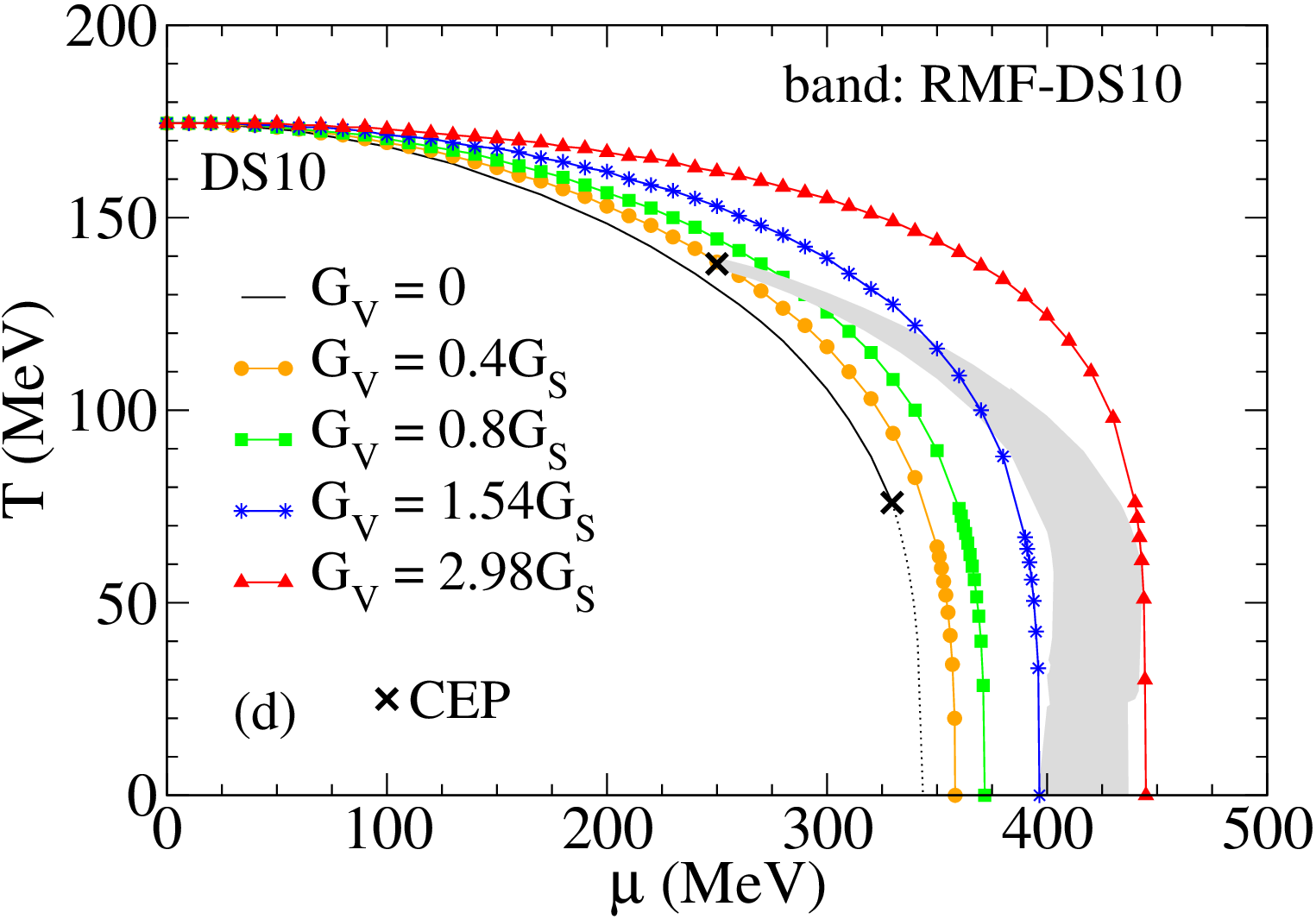}
\caption{(Color online) \mbox{PNJLv} phase diagrams results for different $G_V$
values compared to RMF-PNJL models. The crosses indicate the critical
end-points.}
\label{figs}
\end{figure}
\twocolumngrid

In terms of the ratio $G_V/G_s$, the total range obtained here is \mbox{$1.52
\lesssim G_V/G_s \lesssim 3.2$}. This result suggests that the $G_V$ values
calculated from hadron-quark phase transition curves, are in fact greater
than some used in the literature, namely, $G_V/G_s=0.25$ \cite{ratio1} and
$G_V/G_s=0.5$ \cite{ratio2}. In Ref. \cite{KasPRD11} it was argued that the
range of acceptable values in the non-local \mbox{PNJLv} model are
$0.25<G_V/G_s<0.5$. This range is substantiated by a Fierz transformation of an
effective one-gluon exchange interaction, with $G_V$ depending on the strength
of the $U_A(1)$ anomaly in the two-flavor model. Our values are considerable
above $0.5$, which should not be unexpected, as the vector term in the
\mbox{PNJLv} model, in our fitting procedure, mimics the repulsive short-range
part of the nucleon-nucleon interaction present in the RMF model. The nature of
the repulsive nuclear force is beyond one gluon exchange and acts between
colorless hadron degrees of freedom, that essentially are of non-perturbative
origin, explaining the difference in the values of $G_V$. Consistently, our
values are larger than, for instance, $G_V/G_s=1$ used for the \mbox{PNJLv}
model of Ref. \cite{ratio3}.

It is important to stress that our estimated range for the vector coupling
strength, is compatible with those found recently in Ref.~\cite{ferroni}, given
by \mbox{$4$ GeV$^{-2}\lesssim G_V\lesssim 19$ GeV$^{-2}$}. Actually that range
encompasses ours. In that work, the authors estimated such range through a
completely different way. They used the lattice QCD data of the diagonal and
off-diagonal quark susceptibilities, as input in an effective QCD mean-field
model approach, described by a Lagrangian density of a two-flavor quark system
interacting only via massive vector fields. By doing so, they assume that the
vector part of the QCD interaction can be isolated, and treated in that
approximation. Actually, the authors considered a temperature dependent
vector coupling, $G_V=G_V(T)$. They obtained the range \mbox{$4$
GeV$^{-2}\lesssim G_V(T_c) \lesssim 19$ GeV$^{-2}$} by assuming a transition
temperature of $T_c(\mu=0)=170$~MeV. In our calculations we consider a fixed
vector interaction strength.

For the sake of completeness we also present in Fig. \ref{figs2} the order
parameters $\rho_s$ and $\Phi$, and the quark density, obtained
self-consistently from the \mbox{PNJLv} EOM given by Eqs.
(\ref{qdens})-(\ref{qphi}). All the curves were constructed for the RRW06 model
and for some values of $\mu$. Notice that the dependence of $\rho_s$ and $\Phi$
with temperature becomes smoother when $G_V$ increases. This is consistent
with our findings that the vector interaction in \mbox{PNJLv} models favors the
crossover in the quark phase diagram (see Fig. \ref{figs}). Note also that, the
increase of $G_V$ decreases the quark density for fixed $\mu$ and
$T$ values, as expected from the relation $\rho=(\mu-\tilde{\mu})/2G_V$.

The approaches presented here for the construction of the hadron-quark phase
transitions, can be understood from a qualitatively analysis of the structure of
the PNJL model. It is known that the differences between the NJL and the PNJL
models, are the modification in the Fermi-Dirac distributions and the inclusion
of a Polyakov potential, $\,\mathcal{U}(\Phi,\Phi^*,T)\,$, in the equations of
state. Regarding the statistical distributions, the extreme case in which
$\Phi=\Phi^*=0$ leads, as one can verify from Eq. (\ref{fdmp}) for $G_V=0$, to
$F(k,T,\mu,\Phi=\Phi^*=0)= [e^{3(E-\mu)/T} + 1]^{-1}.$ The same result is
obtained for the anti-quarks distributions by replacing $\mu$ by $-\mu$.

\onecolumngrid
\textcolor{white}{asdfas}
\begin{figure}[!htb]
\hspace{0.4cm}
\includegraphics[scale=0.207]{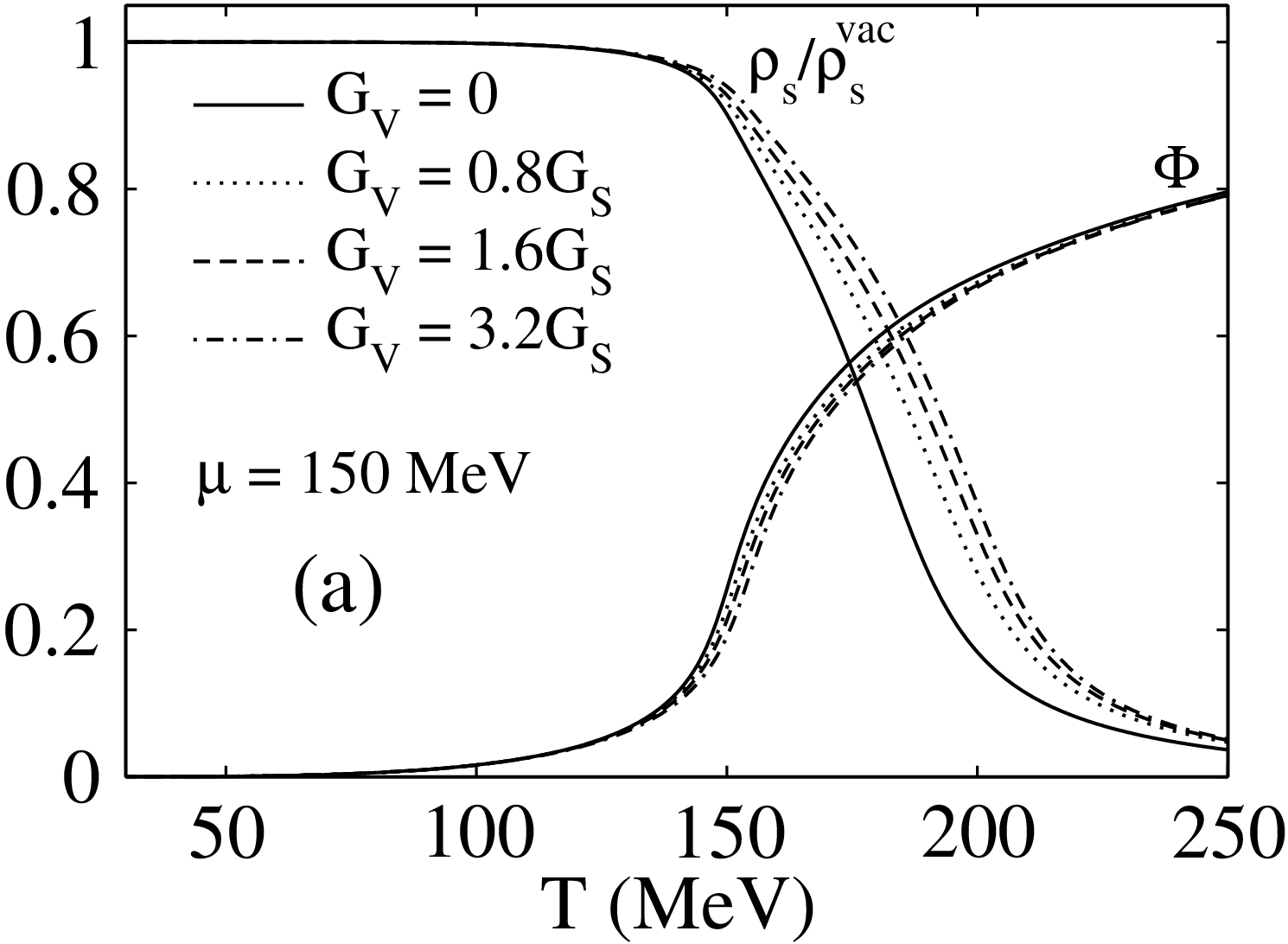}
\includegraphics[scale=0.207]{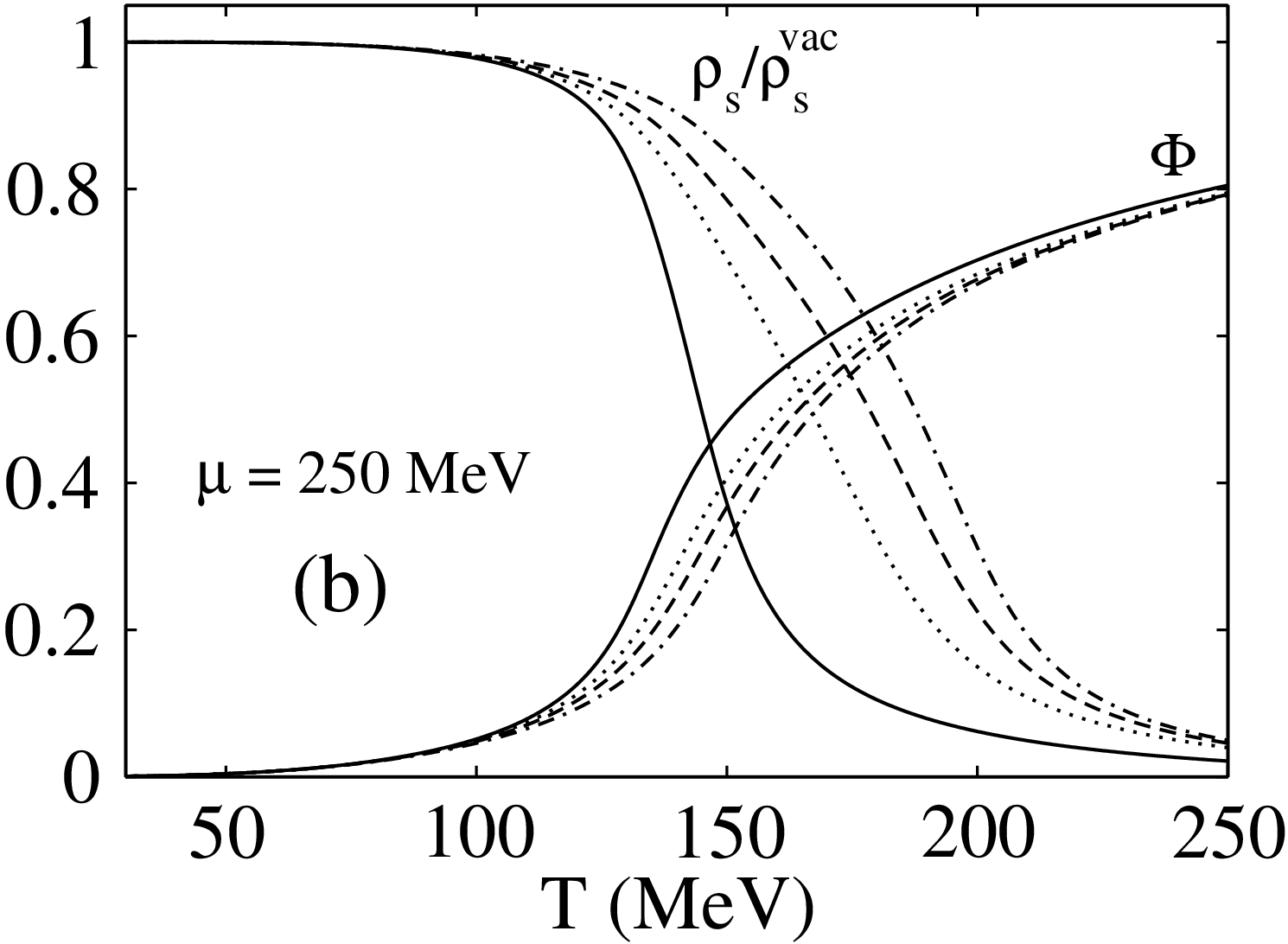}
\includegraphics[scale=0.207]{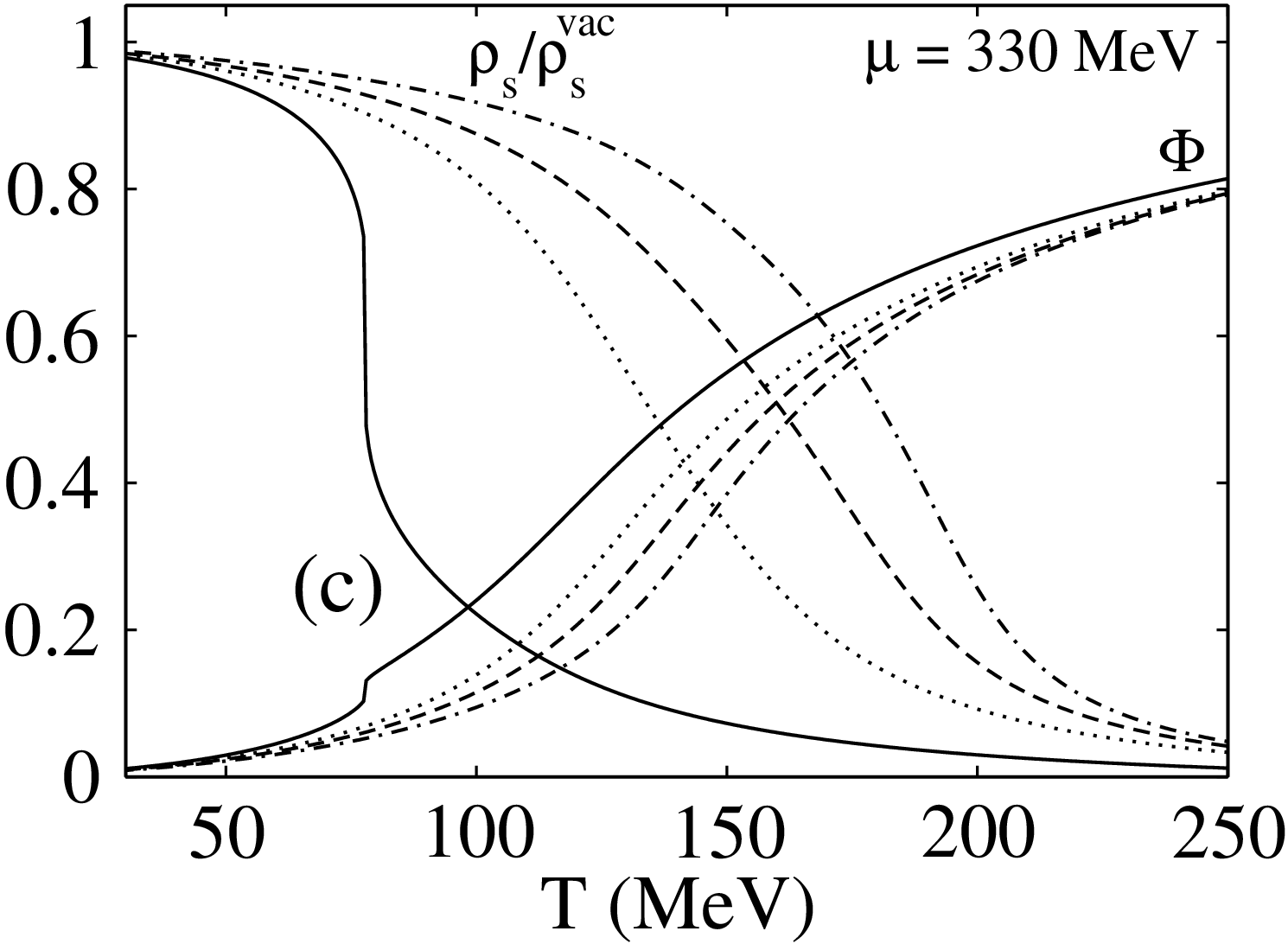}
\end{figure}
\vspace{-1.2cm}
\begin{figure}[!htb]
\includegraphics[scale=0.207]{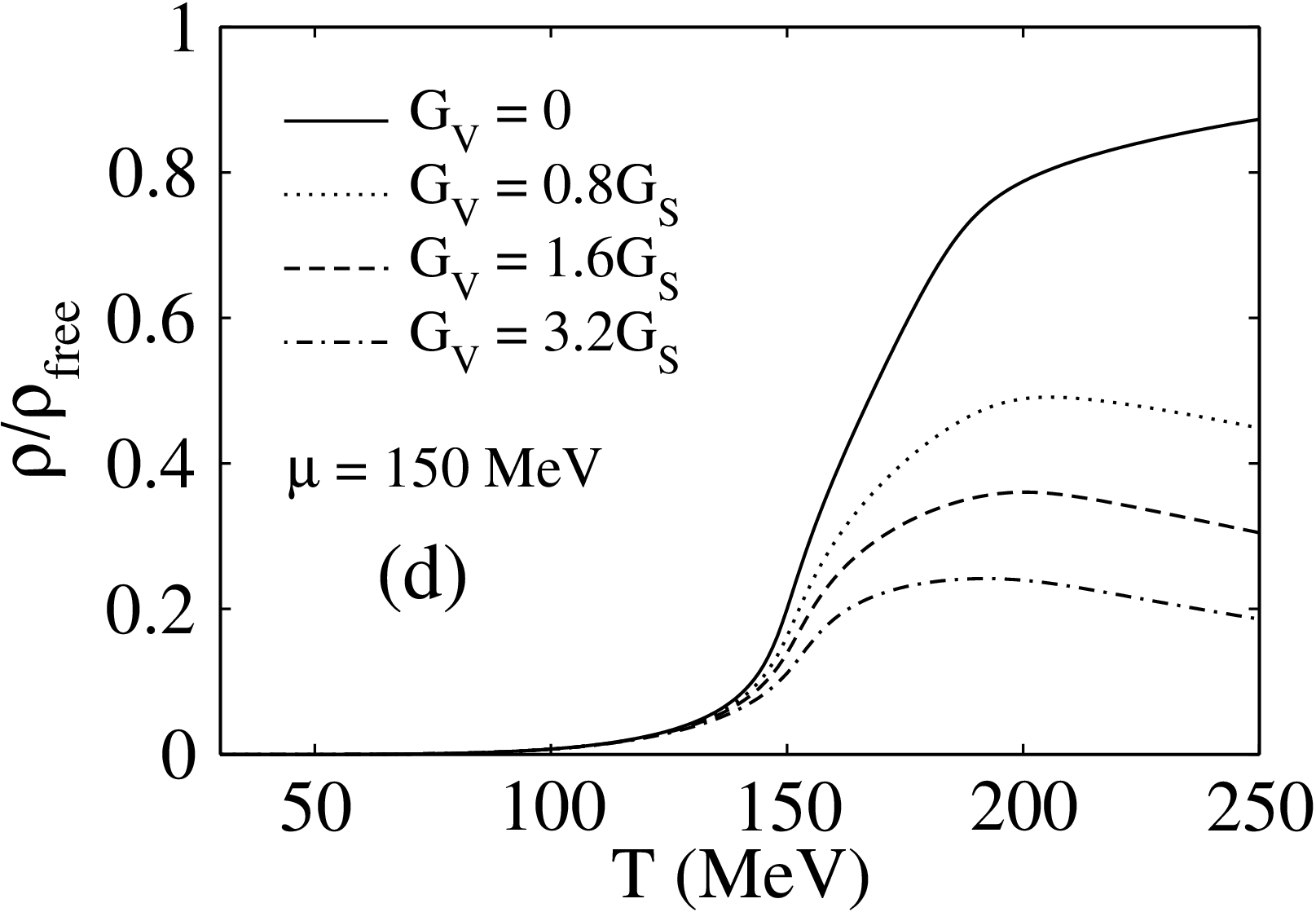}
\includegraphics[scale=0.207]{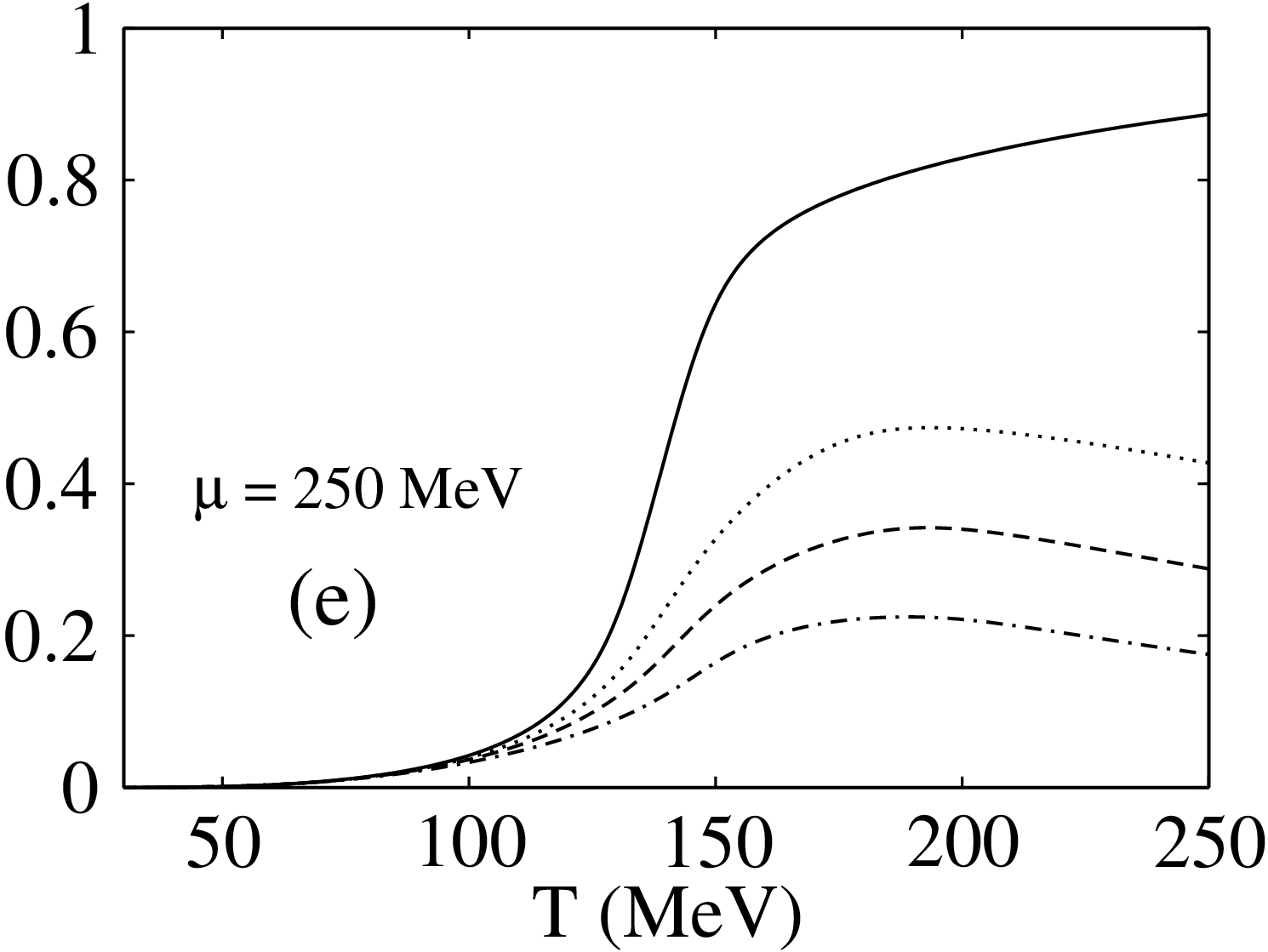}
\includegraphics[scale=0.207]{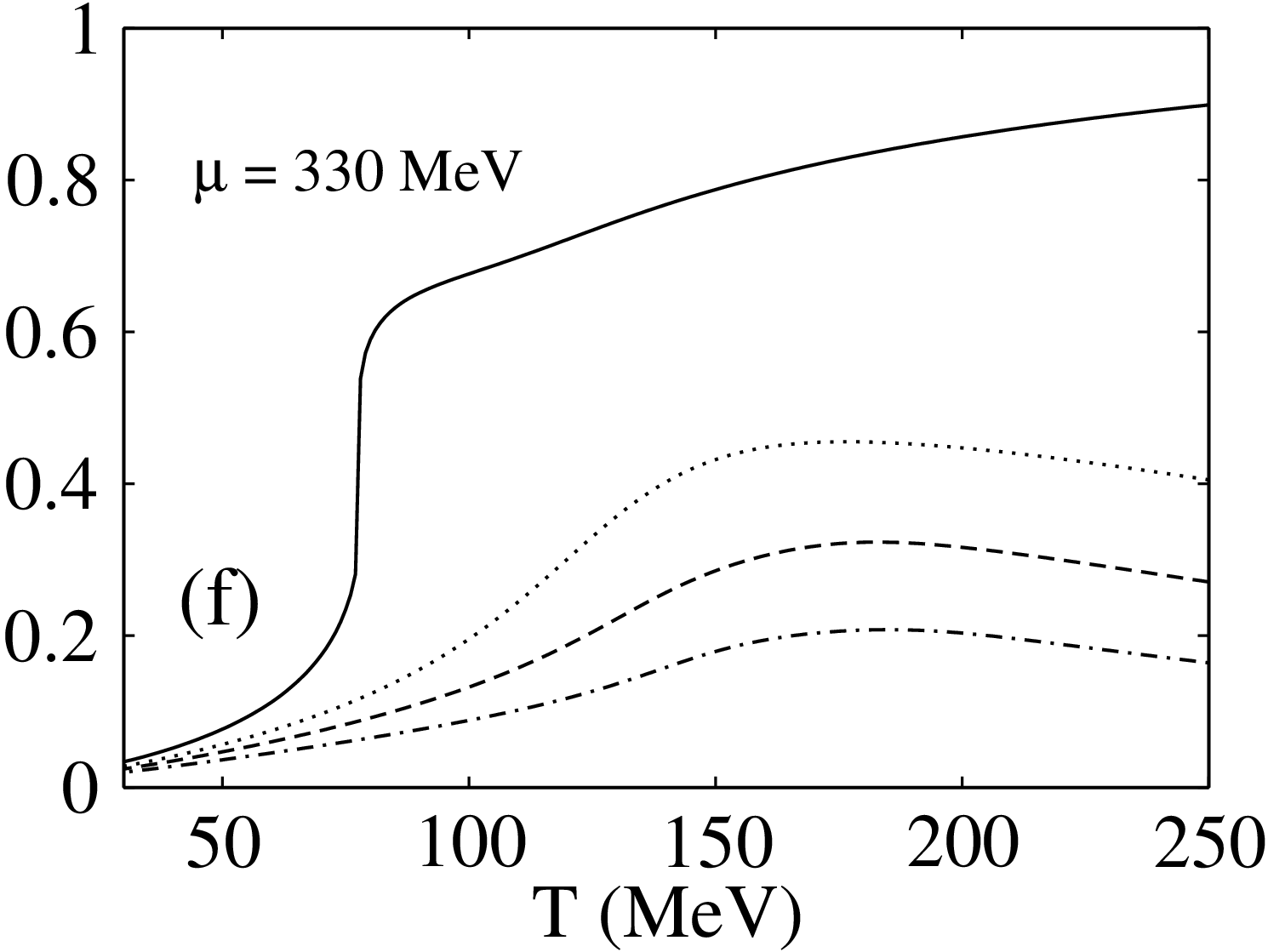}
\caption{(a)-(c): Order parameters versus temperature. (d)-(f): Quark
density normalized by the free massless quark density,
$\rho_{\mbox{\tiny{free}}}=2\mu^3/\pi^2 + 2T^2\mu$, versus temperature. In all
figures the RRW06 parametrization were used.}
\label{figs2}
\end{figure}
\twocolumngrid

Notice that in the case of total confinement, i.e. $\Phi=\Phi^*=0$, the PNJL
model gives rise to the Fermi-Dirac statistics of a three-quark cluster, which
can be seen as a prototype of hadrons. The PNJL model embodies the formation of
hadronic degrees of freedom. From this point of view, the interaction has to
recognize the hadronic formation, and should account for a more realistic
description of the case in which the quarks are totally confined. Therefore, the
interaction between the clusters should also contain the characteristic
repulsion present in the force between hadrons. This feature is qualitatively
built in PNJL models, by including the term $G_V(\bar{\psi}\gamma_\mu\psi)^2$ in
the Lagrangian density (\mbox{PNJLv} model).

In the \mbox{RMF-PNJL} approach of the hadron-quark phase transition
with no vector interaction in the PNJL model, all the repulsion is
restricted to the hadronic sector. What our results reveal is that
the effect of such repulsion in the hadron-quark transition, can be
reproduced in \mbox{PNJLv} models by adjusting the vector interaction
strength. However, we should point out that at high $\mu$ and
low $T$, in the region where the baryonic degrees of freedom
dominate, models formulated only with quark degrees of freedom are
not strictly valid (see e.g. \cite{WeiArxiv12}). The \mbox{PNJLv} model
should also be limited in the same sense, while it still
incorporates the nuclear repulsion close to the hadron-quark phase
transition. As it is known~\cite{fuku2}, by increasing the vector
interaction strength in the \mbox{PNJLv} model the critical end-point (CEP)
disappears, while in the \mbox{RMF-PNJL} model in all cases it still remains. We
stress that the vector interaction strength, in our model, was fitted in a
region of $T\times \mu$ far from the \mbox{RMF-PNJL} critical end-point (see
Fig. \ref{figs}). The present version of the \mbox{PNJLv} model should be
improved to account for a possible CEP. For instance, by including density and
$T$ dependence on the model parameters, in order to incorporate more effects
from QCD. In particular, the recent calculation of the Polyakov quark-meson
model with the functional renormalization group method \cite{HerPLB11}, already
shown that the $\mu$-dependence of $T_0$ affects the QCD phase diagram at finite
$\mu$.

As a last remark, we point out that the construction of the hadron-quark phase
transition is still an open question, from the experimental point of view.
Experiments planned to occur in the new facilities, such as the Facility for
Antiproton and Ion Research (FAIR) \cite{fair} at GSI, and the Nuclotron-based
Ion Collider Facility (NICA) \cite{nica} at the Joint Institute for Nuclear
Research (JINR), will be performed to reach the highly compressed matter
covering thus, as far is possible, the strongly interacting matter phase
diagram. For this reason, it is always important to present and discuss the
different predictions of effective models in the intermediate $T$ and $\mu$
values of the hadron-quark phase diagram, where the lattice QCD calculations
still can not reach.

\acknowledgments
This work was supported by the Brazilian agencies FAPESP and CNPq. M. M. also
acknowledges the support of the FAPESP thematic project 2007/03633-3.


\begin{thebibliography}{99}

\bibitem{lattice} E. Laermann, O. Philipsen, Annu. Rev. Nucl. Part. Sci.
{\bf 53}, 163 (2003); C. R. Allton {\it et al.}, Phys. Rev. D {\bf 71}, 054508
(2005); Z. Fodor, S. D. Katz, and C. Schmidt, J. High Energy Phys. 03,
121 (2007); P. de Forcrand and O. Philipsen, Nucl. Phys. {\bf B673}, 170 (2003).

\bibitem{mit} A. Chodos, R. L. Jaffe, K. Johnson, C. B. Thorn and V. F.
Weisskopf, Phys. Rev. D {\bf 9}, 3471 (1974); A. Chodos, R. L. Jaffe, K.
Johnson, C. B. Thorn, Phys. Rev. D {\bf 10}, 2599 (1974); T. DeGrand, R. L.
Jaffe, K. Johnson, J. Kiskis, Phys. Rev. D {\bf 12}, 2060 (1975).

\bibitem{njl} Y. Nambu, G. Jona-Lasinio, Phys. Rev. {\bf 122}, 345 (1961);
{\bf 124}, 246 (1961); S. P. Klevansky, Rev. Mod. Phys. {\bf 64}, 649 (1992).

\bibitem{buballa} M. Buballa, Phys. Rep. {\bf 407}, 205 (2005); U. Vogl and W.
Weise, Prog. Part. Nucl. Phys. {\bf 27}, 195 (1991); T. Hatsuda and T. Kunihiro,
Phys. Rep. {\bf 247}, 221 (1994).

\bibitem{fuku1} K. Fukushima, Phys. Lett. B {\bf 591}, 277 (2004).

\bibitem{fukureview} K. Fukushima and T. Hatsuda, Rep. Prog. Phys. {\bf 74},
014001 (2011).

\bibitem{kapusta} E. S. Bowman, and J. I. Kapusta, Phys. Rev. C {\bf 79},
015202 (2009).

\bibitem{cratti} C. Ratti and W. Weise, Phys. Rev. D {\bf 70}, 054013 (2004);
I. General, D. G. Dumm, and N. N. Scoccola, Phys. Lett. B {\bf 506}, 267 (2001).

\bibitem{weise4} S. Rossner, C. Ratti, and W. Weise, Phys. Rev. D {\bf 75},
034007 (2007).

\bibitem{fuku2} K. Fukushima, Phys. Rev. D {\bf 77}, 114028 (2008).

\bibitem{outros} K. Fukushima, Phys. Lett. B {\bf 695}, 387 (2011);
C. Sasaki, B. Friman, and K. Redlich, Phys. Rev. D {\bf 75}, 074013 (2007);
H. Abuki, R. Anglani, R. Gatto, G. Nardulli, and M. Ruggieri, Phys. Rev. D
{\bf 78}, 034034 (2008); D. Gomez Dumm, D. B. Blaschke, A. G. Grunfeld, and N.
N. Scoccola, Phys. Rev. D {\bf 78}, 114021 (2008); G. A. Contrera, M. Orsaria,
and N. N. Scoccola, Phys. Rev. D {\bf 82}, 054026 (2010).

\bibitem{plsm} H. Mao, J. Jin, and M. Huang, J. Phys. G {\bf 37}, 035001 (2010).

\bibitem{prd} O. Louren\c co, M. Dutra, A. Delfino and M. Malheiro, Phys. Rev.
D {\bf 84}, 125034 (2011).

\bibitem{ditoro} M. Di Toro, T. Gaitanos, {\it et. al.}, Nucl. Phys. {\bf A775},
102 (2006); M. Di Toro, B. Liu, {\it et. al.}, Phys. Rev. C {\bf 83}, 014911
(2011); G. Y. Shao, M. Di Toro, {\it et. al.}, Phys. Rev. D {\bf 84}, 034028
(2011); G. Y. Shao, M. Di Toro, {\it et. al.}, Phys. Rev. D {\bf 83}, 094033
(2011); B. Liu, M. Di Toro, {\it et. al.}, Eur. Phys. J. A {\bf 47}, 104 (2011).

\bibitem{ciminale} M. Ciminale, R. Gatto, N.D. Ippolito, G. Nardulli,
M. Ruggieri, Phys. Rev. D {\bf 77}, 054023 (2008).

\bibitem{debora} R. Cavagnoli, C. Provid\^encia, and D. P. Menezes, Phys. Rev.
C {\bf 83}, 045201 (2011); M. G. Paoli, D. P. Menezes, Eur. Phys. J. A {\bf 46},
413 (2010).

\bibitem{delfino} A. Delfino, M. Chiapparini, M. E. Bracco, L. Castro, and S.
E. Epsztein, J. Phys. G {\bf 27}, 2251 (2001); A. Delfino, J. B. Silva, M.
Malheiro, M. Chiapparini, and M. E. Bracco, J. Phys. G {\bf 28}, 2249 (2002).

\bibitem{muller} H. M\"uller, Nucl. Phys. {\bf A618}, 349 (1997).

\bibitem{kashiwa} K. Kashiwa, H. Kouno, M. Matsuzaki, and M. Yahiro, Phys.
Lett. B {\bf 662}, 26 (2008).

\bibitem{kashiwa2} K. Kashiwa, H. Kouno, T. Sakaguchi, M. Matsuzaki, and M.
Yahiro, Phys. Lett. B {\bf 647}, 446 (2007); M. Buballa, and M. Oertel, Nucl.
Phys. {\bf A642}, 39c (1998).

\bibitem{pnjlgv} J. Steinheimer, S. Schramm, Phys. Lett. B {\bf 696}, 257
(2011); K. Fukushima, Phys. Rev. D {\bf 78}, 114019 (2008); S.
Carignano, D. Nickel, and M. Buballa, Phys. Rev. D {\bf 82}, 054009
(2010); Y. Sakai, K. Kashiwa, {\it et. al.}, {\bf 78}, 076007
(2008); K. Kashiwa, H. Kouno, and M. Yahiro, Phys. Rev. D {\bf 80},
117901 (2009).


\bibitem{KasPRD11} K. Kashiwa, T. Hell, and W. Weise, Phys. Rev. D {\bf 84}, 056010
(2011).

\bibitem{ratio3}  Y. Sakai, K. Kashiwa, H. Kouno, M. Matsuzaki, M. Yahiro,
Phys. Rev. D {\bf 79}, 096001 (2009).

\bibitem{ferroni} L. Ferroni and V. Koch, Phys. Rev. C {\bf 83}, 045205 (2011).

\bibitem{weise2} C. Ratti, S. Roessner, M.A. Thaler, and W. Weise,
Eur. Phys. J. C {\bf 49}, 213 (2007).

\bibitem{weise1} C. Ratti, M. A. Thaler, and W. Weise, Phys. Rev. D {\bf 73},
014019 (2006).

\bibitem{schramm} V. A. Dexheimer, and S. Schramm, Phys. Rev. C {\bf 81}, 045201
(2010); V. A. Dexheimer, and S. Schramm, Nucl. Phys. {\bf A827}, 579 (2009).

\bibitem{weise6} S. Roessner, T. Hell, C. Ratti, and W. Weise, Nucl. Phys.
{\bf A814}, 118 (2008).

\bibitem{sc1} S. B. R\"uster, V. Werth, M. Buballa, I. A. Shovkovy, and D.
H. Rischke, Phys. Rev. D {\bf 72}, 034004 (2005); D. Blaschke, S. Fredriksson,
H. Grigorian, A. M. Oztas, and F. Sandin, Phys. Rev. D {\bf 72}, 065020 (2005).

\bibitem{sc2}  D. Blaschke, J. Berdermann, and R. Lastowiecki, Prog. Theor.
Phys. Suppl. {\bf 186}, 81 (2010).

\bibitem{stringent} F. Karsch, E. Laermann, and A. Peikert, Nucl. Phys.
{\bf B605}, 579 (2001); F. Karsch, Nucl. Phys. {\bf A698}, 199 (2002); F.
Karsch, Lect. Notes Phys. 583, 209 (2002); O. Kaczmarek and F. Zantow, Phys.
Rev. D {\bf 71}, 114510 (2005).

\bibitem{gshao} G. Y. Shao, Phys. Lett. B {\bf 704}, 343 (2011).

\bibitem{manuel} M. Malheiro, M. Fiolhais, and A. R. Taurines, J. Phys. G
{\bf 29}, 1045 (2003); J. G. Coelho, C. H. Lenzi, M. Malheiro, R. M. Marinho
Jr, C. Provid\^encia and M. Fiolhais, Nucl. Phys. B, Proc. Suppl. {\bf 199}, 325
(2010).

\bibitem{jaziel} J. G. Coelho, C. H. Lenzi, M. Malheiro, R. M. Marinho Jr,
and M. Fiolhais, Int. J. Mod. Phys. D {\bf 19}, 1521 (2010);
C. H. Lenzi, C. V. Flores, G. Lugones, PoS {\bf 046} (2011);
C. H. Lenzi, A. S. Schneider, C. Provid\^encia and R. M. Marinho Jr, Phys. Rev.
C {\bf 82}, 015809 (2010).

\bibitem{ratio1} M. Kitazawa, T. Koide, T. Kunihiro, and Y. Nemoto, Prog. Theor.
Phys. {\bf 108}, 929 (2002).

\bibitem{ratio2} T. Hatsuda, and T. Kunihiro Prog. Theor. Phys. {\bf 74}, 765
(1985).

\bibitem{WeiArxiv12} W. Weise, Prog. Part. Nucl. Phys. {\bf 67}, 299 (2012).

\bibitem{HerPLB11} T. K. Herbst, J. M. Pawlowski, and B.-J. Schaefer,
Phys. Lett. B {\bf 696}, 58 (2011).

\bibitem{fair} C. H\"ohne, K. E. Choi, V. Dobyrn, {\it et. al.}, Nucl. Instr.
and Meth. A {\bf 639}, 294 (2011); P. Senger, T. Galatyuk, {\it et. al.},
J.Phys. G: Nucl. Part. Phys. {\bf 36}, 064037 (2009); S. Chattopadhyay, J. Phys.
G: Nucl. Part. Phys. {\bf 35}, 104027 (2008); J. M. Heuser (CBM Collaboration),
Nucl. Phys. {\bf A830}, 563c (2009); http://www.gsi.de/fair.

\bibitem{nica} A. N. Sissakian, A. S. Sorin, and V. D. Toneev,
arXiv:nucl-th/0608032; http://nica.jinr.ru/.


\end{thebibliography}
\end{document}